\documentclass[final,5p,times,twocolumn]{elsarticle}
\usepackage{hhline}

\journal{Elsevier}  

\begin{document}
\begin{frontmatter}

\title{Three-body calculation of the $1s$ level shift
 in kaonic deuterium}

\author[Bp]{P. Doleschall}
\author[Bp]{J. R\'{e}vai}
\author[Rez]{N.V. Shevchenko\corref{cor1}\ead{shevchenko@ujf.cas.cz}}
\cortext[cor1]{Corresponding author}

\address[Bp]{Wigner Research Center for Physics, RMI,
H-1525 Budapest, P.O.B. 49, Hungary}
\address[Rez]{Nuclear Physics Institute, 25068 \v{R}e\v{z}, Czech Republic}

\date{\today}

\begin{abstract}
The first exact calculation of a three-body hadronic atom was performed.
Kaonic deuterium $1s$ level shift and width were evaluated using
Faddeev-type equations with Coulomb interaction. The obtained exact results
were compared with commonly used approximate approaches.
\end{abstract}


\end{frontmatter}

%\section{Introduction}
%\label{intro}

Kaonic deuterium is very useful exotic atom, which can be
accurately studied experimentally and theoretically and after
comparing the results give us additional information about
antikaon - nucleon interaction.

Interaction of an antikaon with a nucleon is the basis
for investigation of strong quasi-bound states in
antikaonic-nucleus systems, attracted large interest recently.
The most interesting and being intensively studied theoretically
and experimentally is the lightest $K^- pp$ system, see e.g.~\cite{our2014_II}.
At present, the theoretical predictions for binding energies
and widths of the quasibound state differ substantially. The theoretical
results, however, agree that the quasibound state really can exist in
the $K^- pp$ system. The experimental results also differ from each
other; moreover, their binding energies and widths are far from all
theoretical  predictions. Since the question of the possible
existence of the quasibound state in $K^− pp$ system is still
actual, new experiments are being planned and performed by HADES
and LEPS Collaborations, in J-PARC E15 and E27 experiments.

There are two origins of uncertainty for theoretical results.
The first one is different few-body methods, which were used for the
calculations. However, it was shown in~\cite{our2014_II} that even
calculations using the same three-body Faddeev-type equations
give different results in dependence on the $\bar{K}N$ potential
used as the input.

The problem is that available two-body experimental information on
the $\bar{K}N$ interaction is insufficient for construction of a unique
interaction model. In particular, it was shown e.g. in~\cite{NinaKd}
that phenomenological models of the interaction having one or two poles
for the $\Lambda(1405)$ resonance and reproducing all low-energy
experimental data on $K^− p$ scattering and kaonic hydrogen equally well
can be constructed. The same is true for the recently constructed
in~\cite{our2014_I} chirally-motivated model of the interaction.
A way to obtain some additional information about the $\bar{K}N$
interaction is to use it as an input in an accurate few-body calculation
and then compare the theoretical predictions with eventual experimental data.
Kaonic atoms, in contrast to kaonic nuclear states, can be measured
accurately. Kaonic deuterium is the best candidate since energy
shift and width of its $1s$ level can be measured directly and calculated
accurately. It allows a direct comparison of the theoretical predictions
with experimental data on kaonic deuterium, which hopefully
will be obtained in the SIDDHARTA-2 experiment~\cite{SIDD2}.

Characteristics of kaonic deuterium, however, are hard to calculate
accurately. Due to this only approximate formulas such as Deser~\cite{Deser}
or corrected Deser~\cite{corDeser}, connecting $1s$ level shift and width
of an atom with the corresponding scattering length, are used by
experimentalists and some theorists. However, it was shown, for example
in~\cite{ourPRC2009}, that even for two-body system, (anti)kaonic
hydrogen, the formulas give quite large error in comparison with the exact
result. Since no three-body effects can be taken into account by
such a formula, the accuracy of the formulas for kaonic deuterium
should be less.

To the best of our knowledge, the most accurate evaluation of kaonic deuterium
characteristics was done recently in~\cite{NinaKd} and repeated~\cite{our2014_I}
with new $\bar{K}N$ potentials. The two-body calculations using effective
optical $K^- - d$ potential together with Coulomb interaction were performed. 
The potential was constructed in such a way that it reproduces elastic amplitudes
of $K^- d$ scattering obtained from the Faddeev calculation with strong
interactions only.

In this letter we present the results of the first exact calculation of
the three-body atom - kaonic deuterium with no reduction to any effective
two-body problem. The obtained energy of the $1s$ level of the atom is
an exact eigenvalue of the corresponding three-body Hamiltonian with all of
its  interactions taken into account simultaneously. The dynamically exact
results are compared with those of the approximate methods.

%%%%%%%%%%%%%%%%%%%%%%%%%%%%%%%%%%%%%%%
%\section{Formalism}

At present there are powerful methods to solve three-body problems,
especially for the somewhat easier task of finding real or complex
eigenvalues: Faddeev equations (in integral or differential form)
or variational methods based on wave function expansion in coordinate space.
However, just for our case both methods face serious difficulties. In the
Faddeev approach we have the everlasting problem of the long range Coulomb
force, which is even worse for attractive interaction. As for the coordinate
space expansions, the main difficulty lies in the presence of two very
different distance scales, both relevant for the calculated level shift.

The usual description of hadronic atoms is based on a two-body picture:
a negatively charged hadron in the Coulomb field of the nucleus. The strong
interaction with the nuclear few- or many-body system is incorporated into
an absorptive nucleus-hadron interaction. Thus the two interaction types,
forming the system are treated not on equal footing: first, the pure strong
interaction problem is reduced to an effective two-body one, while the Coulomb
force is ``added'' as a second step.

Some years ago a method for simultaneous treatment of short range plus
Coulomb forces in three-body problems was proposed~\cite{Papp1}. 
The method was successfully applied for short range plus repulsive Coulomb
forces (nuclear case)~\cite{Papp2}, and purely Coulomb systems with
attraction and repulsion \cite{Papp3}. The present case of three strongly
interacting hadrons with Coulomb attraction between certain pairs, which is
practically inaccessible by other methods, was not considered. 

The basic idea was to transform the Faddeev integral equations into matrix
form using a special discrete and complete set of Coulomb Sturmian functions
as a basis. Coulomb Sturmian functions in coordinate space have the form
\begin{equation}
 \langle \vec{r} | nlm \rangle = \langle \vec{r} | i\rangle
  = N_{nl} r^l e^{-br} L^{2l+1}_{n} (2br)Y_{lm}(\hat{\vec{r}}),
\end{equation}
were $b$ is a range parameter. The functions $\langle\vec{r}|i\rangle$
are orthogonal with the weight function $1/r$, or they form a bi-orthogonal
and complete set with their counter-parts $\langle\vec{r}|\tilde{i}\rangle$
\begin{equation} 
 \left\langle i  \left| \frac{1}{r} \right| j \right\rangle = \delta_{ij},
 \;
 \langle \vec{r} \, | \tilde{i}\rangle = 
 \frac{1}{r} \, \left\langle \vec{r} \, | i \right\rangle,
 \;
 \langle i|\tilde{j}\rangle = \langle \tilde{i}|j\rangle = \delta_{ij}
\end{equation}

The most remarkable feature of this particular basis set is, that in this
representation the matrix of the $(z-h_c)$ operator, where $h_c$ is
the pure two-body Coulomb Hamiltonian, is tridiagonal. Therefore, if we use
this property for evaluation the matrix elements of the two-body Coulomb Green's
function  $\langle \tilde{i}| g_c(z) | \tilde{j} \rangle$, we get an infinite
tridiagonal set of equations, which can be solved exactly, see~\cite{PappAdd}
and references therein.
The same holds for the matrix elements of the free two-body
Green's function $\langle \tilde{i}| g_0(z) | \tilde{j} \rangle$.

The Noble form of the Faddeev equations \cite{Noble} for the $K^- p n$ three-body
problem, when the Coulomb interaction appears in the Green's functions, reads
\begin{eqnarray}
\Psi_{np} &=& \! \! \left(
  z - H_0 - V^s_{np}(x_{np}) +
     \frac{e^2}{|c_{np} \vec{x}_{np}+\vec{y}_{K^-}|}
 \right)^{-1}  \nonumber \\
 &{}& \times V^s_{np}(x_{np})(\Psi_{nK^-} + \Psi_{pK^-}) \nonumber \\
 \Psi_{nK^-} &=&  \left(
  z - H_0 - V^s_{nK^-}(x_{nK^-}) +
     \frac{e^2}{|c_{nK^-} \vec{x}_{nK^-}+\vec{y}_{p}|}
 \right)^{-1}  \nonumber\\
 &{}& \times V^s_{nK^-}(x_{nK^-})(\Psi_{np} + \Psi_{pK^-}) \label{FN}\\
 \Psi_{pK^-} &=& \left(
  z - H_0 - V^s_{pK^-}(x_{pK^-}) +
     \frac{e^2}{x_{pK^-}}
 \right)^{-1} \nonumber \\
 &{}& \times V^s_{pK^-}(x_{pK^-})(\Psi_{np} + \Psi_{nK^-})\nonumber
\end{eqnarray}
where, as usual for equations of Faddeev type, the total three-body wave function
is separated into three components
\begin{equation}
 \Psi = \Psi_{np}(\vec{x}_{pn},\vec{y}_{K^-}) +
 \Psi_{nK^-}(\vec{x}_{nK^-},\vec{y}_{p}) +
 \Psi_{pK^-}(\vec{x}_{pK^-},\vec{y}_{n}).
\end{equation}
In Eqs.~(\ref{FN}) $V^s$ denote the two-body strong potentials,
$(\vec{x}_{pn},\vec{y}_{K^-})$, $(\vec{x}_{pK^-},\vec{y}_n)$ and
$(\vec{x}_{nK^-},\vec{y}_p)$ are the three sets of Jacobi coordinates,
and $c$ are mass coefficients. The Coulomb interaction between the antikaon
and the proton $-e^2/x_{pK^-}$ is the same in all three equations, but
expressed in different coordinates.

Introducing the shorter notation: $\alpha = (pn,K^-)$,  $(pK^-,n)$,
$(nK^-,p)$ for the partition channels, $(\vec{x}_{\alpha}, \vec{y}_{\alpha})$
for Jacobi coordinates and inserting (approximate) unit operators
into the system of equations
\begin{equation}
 \hat{1}_{\alpha} =
 \sum_{\mu}^{N_{\alpha}} |\mu_{\alpha} \rangle \langle \tilde{\mu}_{\alpha}|
 \sim \hat{1}
\end{equation}
\begin{equation}
 \langle \vec{x}_{\alpha},\vec{y}_{\alpha} | \mu_{\alpha}\rangle  =
 \langle \vec{x}_{\alpha} | i  \rangle\langle \vec{y}_{\alpha} | I \rangle =
 \langle \vec{x}_{\alpha} | nl \rangle\langle \vec{y}_{\alpha} | NL \rangle,
\end{equation}
with three-body  $\mu$ and two-body $iI$ quantum numbers
$\mu = iI= nlNL$, we can write the system of equations
for the unknowns
$X^{\alpha}_{\mu} =\langle \tilde{\mu}_{\alpha} | \Psi_{\alpha} \rangle$
to be solved:
\begin{equation}
\label{syst}
X^{\alpha}_{\mu} =\sum_{(\mu)\gamma\ne\alpha} [G_{\alpha}(z)]_{\mu \mu''}
 (V^s_{\alpha})_{\mu'' \mu'''}
 (M_{\alpha \gamma})_{\mu''' \mu'} X^{\gamma}_{\mu'}.
\end{equation}
The eigenvalue equation is ${\rm Det}(A(z))=0$,
where matrix $A(z) = G(z) V^s M$.

The matrix elements of the overlap matrix between basis functions from
different Jacobi coordinate sets
 $(M_{\alpha \gamma})_{\mu \mu'}  =
  \langle \tilde{\mu}_{\alpha} | \mu'_{\gamma} \rangle$
and of the strong potentials
$(V^s_{\alpha})_{\mu \mu'} =
\langle \mu_{\alpha} | V^s_{\alpha} | \mu'_{\alpha} \rangle$
in Eq.(\ref{syst}) can be calculated numerically.
The remaining matrix elements in the kernel - matrix elements of
the partition Green's functions:
%\begin{equation}
\begin{eqnarray}
&{}& [G_{\alpha}(z)]_{\mu \mu'}  \\
 &=& \left\langle \tilde{\mu}_{\alpha} \left|
 \left(
  z - h_0(x_{\alpha}) - h_0(y_{\alpha}) - V^s_{\alpha}(x_{\alpha}) -
 \frac{e^2}{x_{pK^-}}
 \right)^{-1}
\right| \tilde{\mu'}_{\alpha} \right\rangle \nonumber,
\end{eqnarray}
%\end{equation}
are the basic quantities of the method, and their calculation depends
on $\alpha$.

For $\alpha = (pK^-,n)$  the $G_{\alpha}(z)$ is the
Green's function of two non-interacting subsystems. Therefore, it can be
calculated by a convolution integral along a suitable contour in the complex
energy plane \cite{Bianchi}
\begin{equation}
 G_{\alpha}(z) = \oint g^{sc}_{\alpha}(\epsilon;x_{\alpha})
 g^0_{\alpha}(z-\epsilon;y_{\alpha}) d\epsilon,
\end{equation}
where
\begin{eqnarray}
 g^{sc}_{\alpha}(z;x_{\alpha}) &=& \left(
 z -h_0(x_{\alpha}) - V^{s}_{\alpha} + \frac{e^2}{x_{\alpha}}
 \right)^{-1} \\
 g^{0}_{\alpha}(z;y_{\alpha}) &=& \left(
 z - h_0(y_{\alpha})  \right)^{-1}
\end{eqnarray}
are two-body Green's functions with strong plus Coulomb potential and
free  Green's function, respectively.
While values of $[g^{0}_{\alpha}(z;y_{\alpha})]_{II'}$ can be calculated
using the properties of the Coulomb Sturmian basis, the
matrix elements of $[g^{sc}_{\alpha}(z;x_{\alpha}) ]_{ii'}$
can be found from the matrix resolvent equation
%\begin{equation}
\begin{eqnarray}
\label{ResEq}
 [g^{sc}_{\alpha}(z;x_{\alpha}) ]_{ii'} &=&
 [g^{c}_{\alpha}(z;x_{\alpha}) ]_{ii'} \\
 &+& \sum_{i'',i'''} [g^{c}_{\alpha}(z;x_{\alpha}) ]_{ii''} (V^s_{\alpha})_{i''i'''}
  [g^{sc}_{\alpha}(z;x_{\alpha}) ]_{i'''i'}.
  \nonumber
\end{eqnarray}
  %\end{equation}

For the other two channels $\alpha \ne (pK^-,n)$ an intermediate
step is required. In this case the Coulomb potential is given not
in its ``natural'' coordinates, therefore, we first rewrite it
in the form
%\begin{equation}
\begin{eqnarray}
-\frac{e^2}{|c_{\alpha} \vec{x}_{\alpha}+\vec{y}_{\alpha}|} =
- \frac{e^2}{y_{\alpha}} + \left(
 \frac{e^2}{y_{\alpha}} -
  \frac{e^2}{|c_{\alpha} \vec{x}_{\alpha}+\vec{y}_{\alpha}|}
\right) 
\nonumber \\
 = V^{c,ch}_{\alpha}(y_{\alpha}) + U_{\alpha}(\vec{x}_{\alpha},\vec{y}_{\alpha}).
\end{eqnarray}
%\end{equation}
Here $V^{c,ch}_{\alpha}(y_{\alpha}) $ is the channel Coulomb interaction and
$U_{\alpha}(\vec{x}_{\alpha},\vec{y}_{\alpha})$ is a (short range)
polarization potential, entering in the equation
\begin{equation}
\label{Geq}
 G_{\alpha}(z) = G^{ch}_{\alpha}(z) + G^{ch}_{\alpha}(z) U_{\alpha}
 G_{\alpha}(z).
\end{equation}
for calculation of $G_{\alpha}(z)$. Its matrix elements
$(U_{\alpha})_{\mu\mu'} =
\langle \mu_{\alpha} | U_{\alpha} | \mu'_{\alpha} \rangle$
can be evaluated by numerical integration. The channel three-body Green's function
in Eq.(\ref{Geq}) is defined by
\begin{equation}
 G^{ch}_{\alpha}(z) = \left(
 z - h_0(x_{\alpha}) - h_0(y_{\alpha}) - V^s_{\alpha}(x_{\alpha})
 + \frac{e^2}{y_{\alpha}}
\right)^{-1}
\end{equation}
and, again, it corresponds to non-interacting subsystems, therefore, can be found
by a convolution integral
\begin{equation}
 G^{ch}_{\alpha}(z) = \oint g^{s}_{\alpha}(z;x_{\alpha})
 g^{c,ch}_{\alpha}(z-\epsilon;y_{\alpha}) d\epsilon
\end{equation}
with the two-body Green's functions
\begin{eqnarray}
 g^{s}_{\alpha}(z;x_{\alpha}) &=& \left(
 z -h_0(x_{\alpha}) - V^{s}_{\alpha}
 \right)^{-1}, \\
 g^{c,ch}_{\alpha}(z;y_{\alpha}) &=& \left(
 z - h_0(y_{\alpha}) + \frac{e^2}{y_{\alpha}}  \right)^{-1},
\end{eqnarray}
which can be obtained from resolvent equations similar to Eq.(\ref{ResEq}).

This schematic description of the formalism does not reflects the fact, that
$\bar{K}N$ interaction is isospin dependent and acts in $I=0$ and $I=1$
states. In particle representation it means that the potential
is a $2 \times 2$ matrix:
\begin{equation}
\label{VKN}
 \left(
   \begin{tabular}{cc}
    $V^s_{pK^-,pK^-}$ & $V^s_{pK^-,n\bar{K}^0}$ \\
    $V^s_{n\bar{K}^0,pK^-}$ & $V^s_{n\bar{K}^0,n\bar{K}^0}$
   \end{tabular}
\right),
\end{equation}
therefore, our final equations have 4 Faddeev components, including
$\Psi_{(n\bar{K}^0,n)}$, instead of three. A more detailed description
of the formalism will follow in a subsequent paper.
%------------------------------------------------------------------------------
\begin{table}
\caption{Range $\beta_{\bar{K}N}$ (fm$^{-1}$) and strength $\lambda_{\bar{K}N,I=0}$,
$\lambda_{\bar{K}N,I=1}$ (fm$^{-2}$) parameters of the four complex $\bar{K}N$ potentials
$V_I, V_{II}, V_{III}$, and $V_{IV}$ in isospin representation.}
\label{VKN.tab}
\begin{center}
\begin{tabular}{cccc}
\hline \noalign{\smallskip}
 & $\beta_{\bar{K}N}$ &  $\lambda_{\bar{K}N,I=0}$   &  $\lambda_{\bar{K}N,I=1}$ \\
\noalign{\smallskip} \hline \noalign{\smallskip}
$V_I$ &      $3.0000$  & $-1.7258 -i \, 0.8570$  & $-0.7323 -i \, 0.4201$  \\
$V_{II}$ &  $3.6367$  & $-2.1606 -i \, 0.5937$  & $-1.0998 -i \, 0.4861$  \\
$V_{III}$ & $3.6367$  & $-1.9563 -i \, 0.4534$  & $-0.9761 -i \, 0.3787$  \\
$V_{IV}$ & $2.1978$  & $-0.5669 -i \, 0.2744$  & $-0.1666 -i \, 0.1489$  \\
\noalign{\smallskip} \hline
\end{tabular}
\end{center}
\end{table}
%------------------------------------------------------------------------------

Our aim was to evaluate $1s$ level shift and width of kaonic deuterium
caused by strong interaction between the antikaon and the nucleons.
The reference point $z_0$, from which the energy shift  
$\Delta z = z - z_0$ is measured, is the lowest eigenvalue of the ``dominant''
channel Green's function $G^{ch}_{(pn,K^-)}(z)$. It corresponds
to a deuteron and a kaon ``feeling'' a Coulomb force from the center
of mass of the deuteron. At $z = z_0$ all matrix elements
of $G^{ch}_{(pn,K^-)}(z)$ are singular, the search for
${\rm Det}(A(z)) = 0$ was performed in the vicinity of $z_0$.

In fact, even in the absence of the strong interaction of the kaon
with the nucleons, the presence of the polarization potential
$U_{(pn, K^-)}$ causes a certain real shift of the eigenvalue from
$z_0$ to $z_1$. It reflects the fact, that Coulomb interaction acts
between the antikaon and the proton, not between $K^-$ and $d$.
In principle, the strong shift should be measured from $z_1$ instead
of $z_0$. However, the effect is small, in our case $z_1 - z_0
\approx 10$ eV.
%------------------------------------------------------------------------------
\begin{table*}
\caption{ Kaonic deuterium and hydrogen $1s$ level shifts $\Delta E$ and widths
$\Gamma$ in a form $\Delta (E) - i \, \Gamma/2$ (eV) for the four complex
$\bar{K}N$ potentials. The first and last column show two- and three-body
exact results, respectively, while the approximate Deser, corrected Deser
values and those obtained with optical $K^- - d$ potential are in columns
2,3 and 4, respectively.}
\label{res.tab}
\begin{center}
\begin{tabular}{cccccc}
\hline \noalign{\smallskip}
 & Kaonic hydrogen & 
 \multicolumn{4}{c}{Kaonic deuterium shift} \\
\hhline{~~----}
 & shift &  Deser & Corrected Deser & Complex $V_{K^- - d}$ &  Exact Faddeev\\
\noalign{\smallskip} \hline \noalign{\smallskip}
$V_I$ & $-280 - i \, 268$  & $-723 - i \, 596$  & $-675 - i \, 351$  & $-650 - i \, 434$  & $-641 - i \, 428$\\
$V_{II}$ & $-217 - i \, 292$  & $-732 - i \, 634$  & $-694 - i \, 370$  & $-658 - i \, 460$  & $-646 - i \, 444$\\
$V_{III}$ & $-219 - i \, 293$  & $-837 - i \, 744$  & $-795 - i \, 390$  & $-747 - i \, 517$  & $-732 - i \, 490$\\
$V_{IV}$ & $-280 - i \, 266$  & $-854 - i \, 604$  & $-750 - i \, 310$  & $-740 - i \,422$  & $-736 - i \, 413$\\
\noalign{\smallskip} \hline
\end{tabular}
\end{center}
\end{table*}
%------------------------------------------------------------------------------

The calculation itself demands a rather heavy numerical work with a lot
of small but important technical details. The convergence of the method
depends on the good choice of the range parameters $b$ of the Sturmian
functions. The optimal parameters are different in different partition
channels and are non-equal for the $x$ and $y$ variables. For a good
choice of the $b$-s $30-40$ functions for every variable give an accuracy
of $\sim 0.5 \%$ ($\sim 1-2$ eV). The dimension of the final matrix is
rather large, for $40$ functions in each variable it is about $5000$.

%%%%%%%%%%%%%%%%%%%%%%%%%%%%%%%%%%
%\section{Results and conclusions}

We used four models of $\bar{K}N$ interaction, all are one-term separable
complex potentials with Yamaguchi form factors.
The isospin dependent $I=0$ and $I=1$  potentials
(their parameters are shown in the Table~\ref{VKN.tab})
were transformed
into particle basis, see Eq.(\ref{VKN}). Two potentials reproduce low-energy
characteristics of the $\bar{K}N$ system, obtained from a coupled-channel
chirally motivated model of the interaction from~\cite{our2014_I}: either
the $I=0$ and $I=1$ $\bar{K}N$ scattering lengths or the $K^- p$ scattering
length and the upper pole forming $\Lambda(1405)$ resonance. The other two
models were fitted to the experimental values of $1s$ level shift and width
of kaonic hydrogen, measured by SIDDHARTA Collaboration \cite{Sidd}, and to
the low-energy $K^- p$ cross-sections. As a result, all four potentials give
$1s$ level shift of kaonic hydrogen (presented in the Table~\ref{res.tab})
within or close to the SIDDHARTA data and a reasonable fit to the  elastic
$K^- p \to K^- p$ and charge exchange $K^- p \to \bar{K}^0 n$ cross-sections.

For the $NN$ interaction we took a separable potential, which
reproduces $NN$ scattering lengths, low-energy phase shifts and
deuteron binding energy in $np$ state. 
%Since the present calculation is considered as a first test of the method
%for the description of three-body hadronic atoms and also as a check of
%commonly used approximations, the choice of the potentials
%was not our main concern.

The results of the calculations are shown in the Table~\ref{res.tab}.
The exact (accurate to $\approx 1$ eV) results are compared with commonly
used approximations, mentioned in the introduction. For the approximate
evaluations we used outputs of our Faddeev calculations of low-energy
$K^- d$ scattering without Coulomb interaction. In particular, the corrected
Deser~\cite{corDeser}
\begin{eqnarray}
\label{Deser}
  \Delta E^{cD} - i \, \frac{\Gamma^{cD}}{2} 
      &=& - 2 \alpha^3 \mu_{K^-d}^2 \, a_{K^-d} \\
  \nonumber
  &{}& \times [1 - 2 \alpha \mu_{K^-d} \, a_{K^- d} \, (\ln \alpha - 1)]
\end{eqnarray}
and the original Deser formula~\cite{Deser}, which contains only the first
term in the brackets of the Eq.(\ref{Deser}), use the $K^- d$ scattering
length $a_{K^- d}$, calculated with the corresponding $\bar{K}N$ potentials.
The effective optical $K^- - d$ potential was fitted to the low-energy
$K^- d$ amplitudes from Faddeev calculation and then used in Lippmann-Schwinger
equation together with Coulomb potential in the same way as in~\cite{NinaKd}.
Keeping in mind relative values of deuteron and Bohr radius of kaonic
deuterium, the approximation seemed well grounded.

It is seen from the table that the original Deser formula can be considered
only as a rough estimate. The corrected Deser gives somewhat better real part
of the $1s$ level shift, but has quite large error for the width of the level.
The most accurate approximation is the exact two-body calculation with the
$K^- - d$ effective optical potential supplemented by the Coulomb interaction.

The next step could be a more realistic calculation of the kaonic deuterium atom.
In particular, the formalism should be extended to treat energy dependent
$\bar{K}N$ interaction models. This is necessary for the inherently energy
dependent chirally motivated potentials and for a proper account of the
$\pi \Sigma$ channel via an energy dependent exact optical potential.

\vspace*{3mm}
\noindent
{\bf Acknowledgments.}
The work was supported by the Czech GACR grant P203/12/2126 and the Hungarian
OTKA grant 109462.

%%%%%%%%%%%%%%%%%%%%%%%%%%%%%%%%%%%%%%

\end{document}